\documentclass[preprint,showkeys,aps]{revtex4}
\usepackage{epsf}
\usepackage{dcolumn}
\usepackage{bm}
\everymath{\displaystyle}
\begin{document}
\title{Polarization of spin-1 particles without an anomalous
magnetic moment in a uniform magnetic field}

\author{Alexander J. Silenko}

\affiliation{Institute of Nuclear Problems, Belarusian State
University, Minsk 220030, Belarus}
\date{\today}

\begin {abstract}
The polarization operator projections onto four directions remain
unchanged for spin-1 particles without an anomalous magnetic
moment in a uniform magnetic field. The approximate conservation
of the polarization operator projections onto the horizontal axes
of the cylindrical coordinate system takes place.
\end{abstract}

\maketitle

It is well-known that the projections of the polarization operator
onto the directions both of the constant uniform magnetic field,
$\bm B$, and of the kinetic momentum, $\bm{\pi}=\bm{ p}-e\bm{ A}$
(where $\bm{ p}\equiv -i\nabla$ is the momentum operator and $\bm{
A}$ is the vector potential of the magnetic field), remain
unchanged for Dirac spin-1/2 particles \cite{ST,BKF}. Such
particles do not possess an anomalous magnetic moment (AMM) and
their $g$ factor is $g=2$. It has been found in Ref. \cite{CJP}
that the projection of the polarization operator onto the
direction defined by the operator $\bm\pi\times\bm B$ is also
constant for the Dirac spin-1/2 particles.

In the present work, we prove that similar properties are valid
for spin-1 particles without the AMM. These properties are exact.
We use the designations $[\dots,\dots]$ and $\{\dots,\dots\}$ for
commutators and anticommutators, respectively, and the system of
units $\hbar=c=1$.

The magnetic moment of any spinning particle can be split into
``normal'' and anomalous parts:
$$\mu=\mu_0+\mu', ~~~\mu_0=\frac{eS}{m}, ~~~ \mu'=\mu-\mu_0.$$

One can introduce the $g$ factor as follow:
$$g=\frac{2m\mu}{eS}.$$

As is known, the preferred $g$ factor is equal to 2 for point-like
particles \cite{ren}. Such particles do not possess the AMM. An
example of a point-like spin-1 particle is the W boson. While the
deuteron is a nucleus, its $g$ factor is also close to 2
($g_d=1.714$).

The general Hamiltonian in the Sakata-Taketani (ST) representation
\cite{SaTa} for spin-1 particles interacting with an
electromagnetic field has been obtained by Young and Bludman
\cite{YB}. However, the use of the ST representation does not give
a possibility to obtain a clear semiclassical limit of the
relativistic quantum mechanics. To find such a limit, one should
use the Foldy-Wouthuysen (FW) representation. Properties of this
representation are unique (see Refs. \cite{FW,CMcK,JMP,PRA} and
references therein). The Hamiltonian and all operators are
block-diagonal (diagonal in two spinors). Relations between the
operators in the FW representation are similar to those between
the respective classical quantities. For relativistic particles in
external fields, operators have the same form as in the
nonrelativistic quantum theory. As a result, the FW representation
provides the best possibility of obtaining a meaningful
semiclassical limit of the relativistic quantum mechanics.

The wave functions of spin-1 particles are pseudo-orthogonal,
e.g., their normalization is defined by the relation
$$\int {\Psi^\dag\rho_3\Psi dV}=\int {(\phi^\dag\phi-\chi^\dag\chi)
 dV}=1, $$
where $\Psi =\left(\begin{array}{c} \phi \\ \chi
\end{array}\right)$ is the six-component wave function
(bispinor).

It is convenient to split six-component matrices on the spin
matrices $S_x,S_y,S_z$ and the Pauli matrices those components act
on the upper and lower spinors:
$$\rho_1=\left(\begin{array}{cc}0&1\\1&0\end{array}\right),~~~
\rho_2=\left(\begin{array}{cc}0&-i\\i&0\end{array}\right),~~~
\rho_3=\left(\begin{array}{cc}1&0\\0&-1\end{array}\right).$$

The Hamiltonian for spin-1 particles is pseudo-Hermitian, that is,
it satisfies the conditions
$$ {\cal H}=\rho_3{\cal H}^\dag\rho_3, ~~~{\cal H}^\dag=\rho_3{\cal H}\rho_3.
$$
Even (diagonal) terms of the Hamiltonian are Hermitian and odd
(off-diagonal) terms are anti-Hermitian.

The operator $U$ transforming the wave function to another
representation should be pseudo-unitary:
$$ U^{-1}=\rho_3 U^\dag\rho_3, ~~~U^\dag=\rho_3 U^{-1}\rho_3. $$

The general ST Hamiltonian derived in Ref. \cite{YB} is given by
\begin{equation}\begin{array}{c}
{\cal H}=e\Phi+\rho_3 m+i\rho_2\frac{1}{m}(\bm S\cdot\bm
D)^2\\-(\rho_3+i\rho_2) \frac{1}{2m}(\bm D^2+e\bm S\cdot\bm B)-
(\rho_3-i\rho_2) \frac{e\kappa}{2m}(\bm S\cdot\bm B)\\-
\frac{e\kappa}{2m^2}(1+\rho_1)\biggl[(\bm S\cdot\bm E)(\bm
S\cdot\bm D)-i \bm S\cdot[\bm E\times\bm D]-\bm E\cdot\bm
D\biggr]\\ +\frac{e\kappa}{2m^2}(1-\rho_1)\biggl[(\bm S\cdot\bm
D)(\bm S\cdot\bm E)-i \bm S\cdot[\bm D\times\bm E]-\bm D\cdot\bm
E\biggr],
\end{array} \label{eq15} \end{equation}
where $\bm S$ is the spin matrix, $\bm B$ is the magnetic field
induction, $\kappa={\rm const}$, and $\bm D=\nabla-ie\bm A$. For
spin-1 particles, the polarization operator is equal to
$\bm\Pi=\rho_3\bm S$.

For the considered case of the particle in the uniform magnetic
field, Eq. (\ref{eq15}) reduces to
\begin{equation}  \begin{array}{c}
{\cal H}=\rho_3\left[m+\frac{\bm\pi^2}{2m}
-\frac{e(\kappa+1)}{2m}\bm S\cdot\bm B\right] \\
+i\rho_2\left[\frac{\bm\pi^2}{2m}-\frac{(\bm\pi\cdot \bm
S)^2}{m}+\frac{e(\kappa-1)}{2m}\bm S\cdot\bm B\right],
\end{array} \label{eq16} \end{equation}
where $\bm\pi=-i\bm D=-i\nabla-e\bm A$ is the kinetic momentum
operator.

The above Hamiltonian can be presented in the form
\begin{equation} {\cal H}=\rho_3 {\cal M}+{\cal
O}, ~~~\rho_3 {\cal O}=-{\cal O}\rho_3, \label{eq7} \end{equation}
\begin{equation}  \begin{array}{c}
{\cal M}=m+\frac{\bm\pi^2}{2m}-\frac{e(\kappa+1)}{2m}\bm S\cdot\bm B, \\
{\cal O}=i\rho_2\left[\frac{\bm\pi^2}{2m}-\frac{(\bm\pi\cdot \bm
S)^2}{m}+\frac{e(\kappa-1)}{2m}\bm S\cdot\bm B\right],
\end{array} \label{eqH} \end{equation}
where ${\cal O}$ is the odd operator anticommuting with $\rho_3$.
The connection between the factors $\kappa$ and $g$ is given by
$g=\kappa+1$ \cite{YB}.

The operators ${\cal M}$ and ${\cal O}$ commute only when
$\kappa=1$ ($g=2$). The relation $[{\cal M},{\cal O}]=0$ is the
sufficient condition of the exact FW transformation \cite{PRA}.
Therefore, the value $g=2$ is prominent not only in the field
theory but also in the quantum mechanics of spin-1 particles. In
this case, the exact FW Hamiltonian is equal to \cite{PRA} $$
{\cal H}_{FW}=\rho_3\sqrt{{\cal M}^2+{\cal O}^2}
$$ or
\begin{equation} {\cal H}_{FW}=\rho_3\sqrt{m^2+\bm
\pi^{2}-2e\bm{S}\cdot\bm B}.  \label{eq1} \end{equation}

This equation is similar to the corresponding one for Dirac
particles ($g=2$) \cite{C}:
\begin{equation} {\cal H}_{FW}=\rho_3\sqrt{m^2+\bm
\pi^{2}-e\bm{\sigma}\cdot\bm B}, \label{eqC} \end{equation} where
$\bm{\sigma}$ is the Pauli matrix. The similarity between Eqs.
(\ref{eq1}) and (\ref{eqC}) results in analogous properties of
spin-1/2 and spin-1 particles interacting with the uniform
magnetic field.

Hamiltonian (\ref{eq1}) commutes with the operator
$\bm{\Pi}\cdot\bm B$. As $\bm{\Pi}\cdot\bm B=\Pi_B B,~ \bm B={\rm
const}$,  the commutator of the operators ${\cal H}_{FW}$ and
$\Pi_B$ equals zero: $[{\cal H}_{FW},\Pi_B]=0$. This ensures
retaining the polarization operator projection onto the magnetic
field direction $\Pi_B$. The relations
$$[\bm\pi^2,\bm\pi]=2ie\bm B\times\bm\pi, ~~~
[\bm{S}\cdot\bm B,\bm{S}\cdot\bm\pi]=i\bm{S}\cdot(\bm
B\times\bm\pi)$$ result in the commutation of the FW Hamiltonian
with the operator $\bm{\Pi}\cdot\bm\pi$. Since
$\bm{\Pi}\cdot\bm\pi=\Pi_l |\bm \pi|$ and $[{\cal
H}_{FW},|\bm\pi|]\equiv[{\cal H}_{FW},\sqrt{\bm\pi^2}]=0$, so
\begin{equation} [{\cal H}_{FW},\Pi_l]=0,
\label{eql}
\end{equation}
where the operator $\Pi_l$ defines the longitudinal projection of
the polarization operator. Therefore, this projection also remains
unchanged.

The longitudinal direction is defined by the operator $\bm\pi$.
However, the unit operator fixing this direction cannot be given
by
\begin{equation} \bm l=\frac12\left(\bm\pi\frac{1}{|\bm\pi|}+
\frac{1}{|\bm\pi|}\bm\pi\right),  \label{equl}
\end{equation}
because $\bm l^2\equiv\bm l\cdot\bm l\neq1$. The square of the
operator $\bm l$ is equal to
$$\begin{array}{c}
\bm l^2=\frac14\left(1+\left\{\frac{1}{|\bm\pi|},
\bm\pi\frac{1}{|\bm\pi|}\bm\pi\right\}+\bm\pi\frac{1}{{\bm\pi}^2}\bm\pi
\right)\\
=1-\frac18\left(\left\{\frac{1}{|\bm\pi|},\left[
\bm\pi,\left[\bm\pi,\frac{1}{|\bm\pi|}\right]\right]\right\}+\left[
\bm\pi,\left[\bm\pi,\frac{1}{{\bm\pi}^2}\right]\right]\right).
\end{array} $$
The double commutators can be calculated with Eqs. (24),(25) from
Ref. \cite{JMP}. When we suppose the parameter $|e|B/{\bm\pi}^2$
to be small with respect to 1, the approximate expression for the
square of the operator $\bm l$ is
\begin{equation} \bm l^2=1+\frac{3e^2B^2}{4{\bm\pi}^4}-
\frac{7e^2(\bm\pi\cdot\bm B)^2}{4{\bm\pi}^6}. \label{lsapp}
\end{equation}

The commutation of the operators ${\cal H}_{FW}$ and
$\bm\Pi\cdot(\bm\pi\times\bm B)$ is proved in a similar way. As
$$\begin{array}{c}\left[\bm \pi^2,(\bm\Pi\cdot[\bm\pi\times\bm B])\right]=
[\bm \pi^2,\bm\pi]\cdot [\bm B\times\bm\Pi]\\=2ie[\bm
B\times\bm\pi]\cdot [\bm B\times\bm\Pi],\\
\left[(\bm{S}\cdot\bm B),(\bm\Pi\cdot[\bm\pi\times\bm B])\right]=
\rho_3\left[(\bm{S}\cdot\bm B),(\bm S\cdot[\bm\pi\times \bm
B])\right]\\=i\rho_3\bm{ S}\cdot\left[\bm B\times[\bm\pi\times\bm
B]\right]=i[\bm B\times\bm\pi]\cdot [\bm
B\times\bm\Pi],\end{array}$$ the operator under the radical sign
in Eq. (\ref{eq1}) commutes with $\bm\Pi\cdot(\bm\pi\times\bm B)$,
and, hence, the Hamiltonian ${\cal H}_{FW}$ also commutes with
this operator: $\left[{\cal H}_{FW}, (\bm\Pi\cdot[\bm\pi\times\bm
B])\right]=0$.

   It is easy to prove \cite{CJP} that the Hamilton operator commutes with
 $$|\bm\pi\times\bm B|\equiv\sqrt{(\bm\pi\times\bm B)^2}=\sqrt{\bm
 \pi^2B^2-(\bm \pi\cdot\bm B)^2}.$$

As $[\pi_i,\pi_j]=iee_{ijk}B_k$ ($e_{ijk}$ is the antisymmetric
unit pseudotensor), so the operator $\bm\pi\cdot\bm B$ commutes
with any projection of the kinetic momentum operator $\bm\pi$.
Since
$$[{\cal H},\bm\pi^2]=0, ~~~
[\bm\pi^2,(\bm\pi\cdot\bm B)^2]= \left\{[\bm\pi^2,(\bm \pi\cdot\bm
B)],(\bm \pi\cdot\bm B)\right\}=0,$$ the operator under the
radical sign in Eq. (\ref{eq1}) commutes with $|\bm\pi\times\bm
B|$. Whence it follows that $\left[{\cal H}_{FW},|\bm\pi\times\bm
B| \right]=0$ \cite{CJP}. As $\bm\Pi\cdot(\bm\pi\times\bm
B)=\Pi_{t}|\bm\pi\times\bm B|$, the polarization operator
projection onto the transversal direction $\bm\pi\times\bm B$
commutes with the Hamiltonian:
\begin{equation} [{\cal H}_{FW},\Pi_t]=0. 
\label{eqt} \end{equation} As a result, the transversal projection
of the polarization operator also remains unchanged at the motion
of the particle.

We can similarly prove that
$$\begin{array}{c}\left[{\cal H}_{FW},\left(\bm\Pi\cdot[\bm B\times(\bm\pi\times\bm
B)]\right)\right]=0, \\ \left[{\cal H}_{FW},|\bm
B\times(\bm\pi\times\bm B)| \right]=0.\end{array}$$ These
relations result in the conservation of the polarization operator
projection onto the direction $\bm B\times(\bm\pi\times\bm B)$:
\begin{equation} [{\cal H}_{FW},\Pi_{B\pi B}]=0,
\label{eqBpB}
\end{equation}
where $\Pi_{B\pi B}$ is defined by $\bm\Pi\cdot[\bm
B\times(\bm\pi\times\bm B)]=\Pi_{B\pi B}|\bm
B\times(\bm\pi\times\bm B)|$. The same property is also valid for
spin-1/2 particles.

The problem of conservation of the polarization operator
projections onto the axes of the cylindrical coordinate system is
rather important \cite{CJP}. If the $z$ axis in parallel to the
field direction ($\bm B=B\bm e_z$), then the polarization operator
projections onto the directions of vectors $\bm e_\rho$ and $\bm
e_\phi$ have the form
$$\begin{array}{c}
\Pi_\rho=\bm\Pi\cdot\bm e_\rho=\Pi_x\cos{\phi}+\Pi_y\sin{\phi},\\
\Pi_{\phi}=\bm\Pi\cdot\bm e_\phi=-\Pi_x\sin{\phi}+\Pi_y\cos{\phi}.
\end{array}$$

These projections would be conserved at condition that
\begin{equation}
[{\cal H},\Pi_\rho]=0, ~~~ [{\cal H},\Pi_\phi]=0, \label{eq3}
\end{equation} or
\begin{equation}\begin{array}{c}
[{\cal F},\Pi_\rho]=0, ~~~ [{\cal F},\Pi_\phi]=0, ~~~ {\cal
F}=m^2+{\bm\pi}^{2}-2e\bm{S}\cdot\bm B.
\end{array}\label{eq4}\end{equation}

   Since
$$\bm{S}\cdot\bm B=S_zB,~~~[S_z,\Pi_\rho]=i\Pi_\phi,~~~
[S_z,\Pi_\phi]=-i\Pi_\rho,$$ the commutators in Eq. (\ref{eq4})
are equal to
\begin{equation}
\begin{array}{c}
\left[{\cal F},\Pi_\rho\right]=-i\Biggl(\biggl\{\pi_\phi,
\frac{\Pi_\phi}{\rho}\biggr\}+2eB\Pi_\phi\Biggr), \\
\left[{\cal F},\Pi_\phi\right] =i\Biggl(\biggl\{\pi_\phi,
\frac{\Pi_\rho}{\rho}\biggr\}+2eB\Pi_\rho\Biggr),   \\
\pi_\phi=\frac12\biggl(-\{\pi_x,\sin{\phi}\}+\{\pi_y,\cos{\phi}\}\biggr).
\end{array} \label{eq5}
\end{equation}

Formulae (\ref{eq5}) which are similar to the corresponding
equations for spin-1/2 particles \cite{CJP} show that conditions
(\ref{eq3}),(\ref{eq4}) cannot be satisfied in the general case.
However, these conditions are approximately satisfied when a
particle motion can be semiclassically described. Semiclassical
description is possible, if the orbital angular moment of
particles is large enough:
$$L=r|P_\phi|\gg 1, $$
where $r$ is the radius of the circular orbit, and $P_\phi$ is the
projection of the classical momentum of particles. In this case,
the radius $r$ is defined by
$$r=-\frac {P_\phi}{eB}$$
and the commutators of the operator $\pi_\phi$ with the coordinate
operators are negligible (see Refs. \cite{BKF,CJP,JETP}).

   The operators $\pi_\phi$ and $\rho$ are defined on the class of functions
that are the eigenfunctions of ${\cal H}$. With allowance for the
semiclassical nature of the motion and the possibility to ignore
the noncommutativity of the operator $\pi_\phi$ with the
coordinate operators, we can therefore replace the operators
$\pi_\phi$ and $\rho$ by the values $P_\phi$ and $r$,
respectively. Thus, conditions (\ref{eq3}),(\ref{eq4}) are
approximately satisfied.

Therefore, the polarization operator projections onto the radial
and azimuthal directions of the axes of the cylindrical coordinate
system, $\Pi_\rho$ and $\Pi_\phi$, are approximately conserved,
when the orbital angular moment of particles is large with respect
to 1. The spin-1/2 particles possess the same property \cite{CJP}.

It is proved that the polarization operator projections onto the
vertical, longitudinal, transversal directions and the direction
orthogonal to the vertical and transversal ones are conserved for
spin-1 particles without the AMM in a uniform magnetic field.
Spin-1/2 particles without the AMM possess the same property. The
conservation of the polarization operator projections onto the
horizontal axes of the cylindrical coordinate system is
approximate.

\end{document}